\documentstyle[12pt,aasms4]{article}

\begin{document}

\title{3rd Interplanetary Network Localization, Time History,
Fluence, Peak Flux, and Distance Lower Limit of the February 28, 1997 Gamma--Ray Burst}
\author{K. Hurley}
\affil{University of California, Berkeley, Space Sciences Laboratory,
Berkeley, CA 94720-7450}
\authoremail{khurley@sunspot.ssl.berkeley.edu}
\author{E. Costa, M. Feroci}
\affil{Istituto Astrofisica Spaziale, CNR, Frascati, Italy}
\author{F. Frontera}
\affil{Universita di Ferrara and ITESRE, Bologna, Italy}
\author{T. Cline}
\affil{NASA Goddard Space Flight Center, Code 661, Greenbelt, MD 20771}
\author{D. Dal Fiume, M. Orlandini}
\affil{ITESRE, Bologna, Italy}
\author{M. Boer}
\affil{CESR, BP 4346, 31029 Toulouse Cedex, France}
\author{E. Mazets, R. Aptekar, S. Golenetskii, M. Terekhov}
\affil{Ioffe Physico-Technical Institute, St. Petersburg, 194021 Russia}

\begin{abstract}
The gamma--ray burst of 1997 February 28 was localized using the arrival--time analysis
method with the Ulysses, BeppoSAX, and WIND spacecraft.  The result is a $\pm31.5 \arcsec
 (3\sigma)$ wide annulus of
possible arrival directions which intersects both the position of the burst determined
independently by the SAX Wide Field Camera, and the position of a fading X-ray
source detected by the SAX focussing X--ray telescopes, and
reduces these source location areas by factors of 7 and 1.5 respectively.  
The combination of the annulus and the SAX locations, a
0.76 square arcminute error box, is consistent
with that of an optical transient source and an extended object, possibly a galaxy.  We also present
the time history, peak flux, and fluence of this event, and derive a model-independent lower
limit to the source distance of $\approx$11000 AU.
\end{abstract}

\keywords{gamma rays: bursts}

\section{Introduction}

The gamma--ray burst (GRB) of 1997 February 28 is the first for which
a fading X--ray source (Costa et al. 1997a; Costa et al. 1997b; Yoshida et al. 1997) and an optical transient (Groot et al.
1997a; van Paradijs et al. 1997) have been reported.  
The position of the optical transient is consistent
with that of an extended object which may be a galaxy (Groot et al. 1997b; Sahu et al. 1997), leading to the possibility that this
may be the first observation of a counterpart to a classical GRB.  The probability
that the GRB, the fading X--ray source, the optical transient, and the extended source/galaxy are
related is in part a function of the position and size of the GRB error box.
In this paper, we present the annulus of possible arrival directions derived by
triangulation between the Ulysses, BeppoSAX, and WIND spacecraft, as well as the
error box formed by the intersection of the annulus with the independently determined BeppoSAX burst
positions.  We also present the time history, peak flux, and fluence
of this burst, and derive a lower limit to its distance based only on the curvature
of the wavefront.

\section{Observations}

GRB 970228 was observed by the Ulysses GRB (Hurley et al. 1992), WIND KONUS 
(Aptekar et al. 1995), and BeppoSAX Wide Field Camera (WFC: Jager et al., 1997) and
Gamma--Ray Burst Monitor (GRBM: this is part of the Phoswich Detection System -
see Frontera et al. 1997a; Costa et al. 1997c; Pamini et al. 1990) 
experiments.  The Ulysses and GRBM light curves
are shown in figure 1.  Ulysses, in heliocentric orbit, was 1988 light-seconds
from Earth; WIND, at the first Lagrange point, was 4.6 light-seconds from Earth,
and BeppoSAX is in low Earth orbit.  For the purposes of GRB triangulation, this
means that the Ulysses/SAX and Ulysses/WIND baselines are practically identical; 
the two independent annuli of burst positions are consistent with one another, but
redundant.

\section{Position}

Figure 2 shows the Ulysses/SAX GRBM triangulation annulus, the SAX WFC position 
(Costa et al. 1997d; Frontera et al. 1997b),
and the position of the fading X--ray source determined by the SAX Low and
Medium Energy Concentrator/
Spectrometers (LECS/MECS: Boella et al. 1997; Parmar et al. 1997).  The position of the extended source/fading optical source is also
indicated.  This annulus is consistent with, but smaller than those previously determined (Hurley et al.
1997a, 1997b; Cline et al. 1997).  Table 1 gives the details of the Ulysses/SAX GRBM and Ulysses/WIND KONUS
annuli, and table 2 gives the intersection points between the Ulysses/SAX GRBM annulus,
and the SAX WFC and LECS/MECS error circles.

The region defined by the SAX WFC error circle and the Ulysses/SAX annulus
has an area of approximately 4 square arcminutes, or 0.14 times the area of the 
SAX WFC error circle; the region defined by the SAX LECS/MECS error circle and the
Ulysses/SAX annulus has an area of approximately 1.5 square arcminutes, or 0.67
times the area of the SAX LECS/MECS error circle.  The combination of the two error
circles and the annulus defines a region with an area of approximately 0.76 square
arcminutes.  Finally, the extended source possibly associated with the optical transient
lies approximately 22.5 \arcsec from the center line of the Ulysses/SAX annulus.

\section{Fluence and Peak Fluxes}

When the distance of the extended source is eventually determined, the
total energy and peak luminosity of this burst will be important parameters
for any models.  Although the Ulysses spectral memory triggered late in the
burst and resulted in too few counts for an 
accurate determination of the energy spectrum, we can calculate the
fluence and peak flux of this burst under the assumption of a power law spectrum
with index -1.07, which is consistent with the spectral data that we have.
For convenience, we have extrapolated these numbers to an
energy range, and have used time intervals, which allow a comparison with cataloged BATSE bursts
(Meegan et al. 1996).
The 25--100 keV fluence was approximately $\rm4.3\times10^{-6} erg\, cm^{-2}$.
The 25--100 keV peak fluxes were approximately $\rm1.4\times10^{-6} erg\, cm^{-2}\,
s^{-1}$ over 0.0625 s and $\rm1.2\times10^{-6} erg\, cm^{-2}\, s^{-1}$ over 0.25
s.

More accurate spectral measurements were made by the KONUS experiments aboard
both the WIND and Kosmos-2326 spacecraft over the 13 keV--10 MeV range.  An optically
thin thermal bremsstrahlung fit (kT=150 keV) gives 25-100 keV fluences and peak
fluxes $\rm5.0\times10^{-6} erg\, cm^{-2}$ and $\rm1.9\times10^{-6} erg\, cm^{-2}\, s^{-1}$,
respectively.  The latter was derived by scaling the average spectrum over
$\rm \sim4 s$ to the 16 ms resolution time history.

The SAX GRBM recorded $\rm2 \times10^{-6} erg\, cm^{-2}$, 40--100 keV,
$\rm8\times10^{-6} erg\, cm^{-2}$, 100--700 keV, and a peak flux
$\rm4 \times10^{-6} erg\, cm^{-2}\, s^{-1}$ (over 1 s, 40--600 keV).

\section{Distance lower limit}

Virtually all distance estimates for GRB's involve models of matter distribution
(e.g. in the Oort cloud, the galactic halo, or at cosmological distances).  It
is therefore interesting to consider model-independent distance limits.  If we
assume that some source within the annulus (say the optical transient)
 is associated with this GRB, we can derive
a distance lower limit from the wavefront curvature.  To do this, we imagine
that the source is at a very large distance from Earth, along the Earth--source vector,
and allow the distance to decrease.  For each value of the assumed distance,
we calculate the difference in arrival times at the Ulysses and SAX satellites and
compare this with the arrival time difference for a source at infinity (i.e. a plane wave).  When
these two times differ by an amount which corresponds to the measured timing
uncertainty in the comparison of the two time histories in figure 1 (300 ms), we have a
source distance lower limit.  In this case, that limit is $\approx$11000 AU
or about 0.05 pc; a refined analysis, currently in progress, may raise this
limit.

\section{Conclusion}

The IPN localization for GRB970228 is consistent with the SAX error circles
and the position of the optical transient and extended source.
SAX WFC localizations of GRBs provide a strong confirmation of IPN accuracies.
Consistency between IPN annuli and SAX error circles are in effect a rigorous
``end--to--end'' test of the spacecraft timing and the software used to derive
burst positions by triangulation.   In the case of GRB970228, this consistency
demonstrates that IPN uncertainties cannot exceed approximately $\pm 3\arcmin$,
the radius of the WFC error circle.  A similar result was obtained for GRB
970111 (Hurley et al. 1997c).  Assuming that the fading X--ray source is
associated with the GRB, the uncertainty becomes approximately  $\pm 1\arcmin$,
and further assuming that the optical transient is associated with the GRB leads
to a maximum uncertainty of approximately $\pm 30\arcsec$.  

\acknowledgments

KH acknowledges support for Ulysses operations under JPL Contract 958056.  On the
Russian side, this work was partially supported by RFBR grant \# 96-02-16860 and
CRDF grant \# RP1-236.  EC, MF, FF, DDF, and MO acknowledge the Agenzia Spaziale
Italiana (ASI) for support of BeppoSAX operations.  The BeppoSAX satellite is
a joint Italian and Dutch project, involving teams at SRON (WFC instrument) and 
at the Space Science Dept. of ESA (LECS instrument), in addition to those in
Italy.  We are grateful to G. Share for prompting us to consider distance lower
limits, and to J. McTiernan for Ulysses spectral analysis.

\begin{figure}
\figurenum{1}
\epsscale{.75}
\plotone{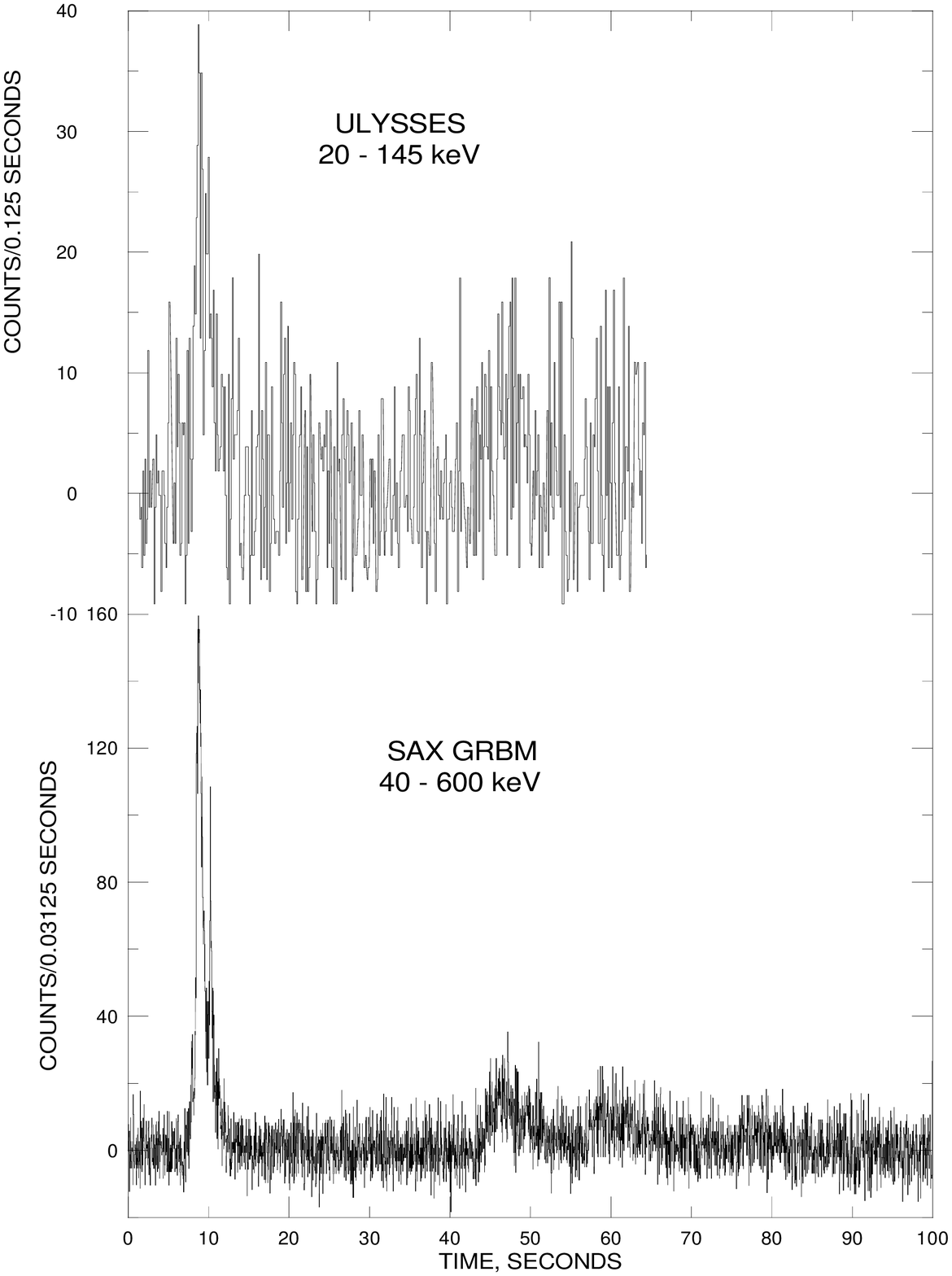}
\caption{Background-subtracted time histories of GRB970228 as observed by Ulysses GRB and
SAX GRBM}
\end{figure}

\begin{figure}
\figurenum{2}
\epsscale{}
\plotone{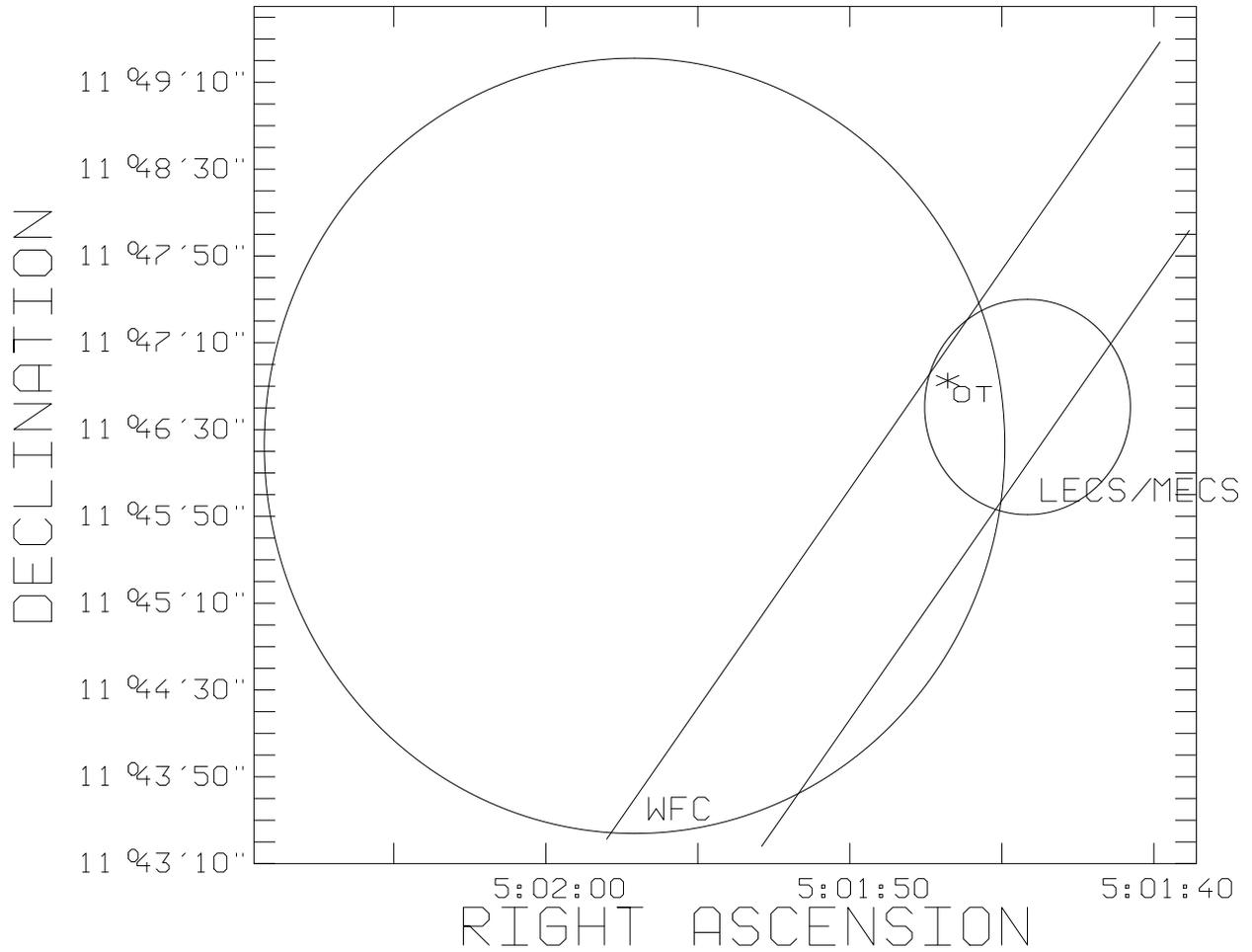}
\caption{SAX WFC, SAX LECS/MECS, and IPN annulus for GRB970228.  For clarity,
only the Ulysses/SAX annulus is indicated; the Ulysses/KONUS annulus is
virtually identical.  The position of
the optical transient (OT) is also indicated.}
\end{figure}

\begin{deluxetable}{ccccc}
\footnotesize
\tablecaption{IPN Annuli}
\tablewidth{0pt}
\tablehead{
\colhead{Spacecraft} & \colhead{Annulus R.A. (2000)} & \colhead{Annulus Decl. (2000)} & 
\colhead{Annulus radius} & \colhead{Annulus halfwidth\tablenotemark{a}}
}
\startdata
Ulysses/SAX & 11\fh 03\fm 47\fs & 36\arcdeg 31\arcmin 31\arcsec & 83.4228\arcdeg & 31.5\arcsec \nl
Ulysses/WIND & 11\fh 03\fm 47\fs & 36\arcdeg 27\arcmin 46\arcsec & 83.4326\arcdeg & 35.7\arcsec \nl
\enddata
\tablenotetext{a}{3$\sigma$ confidence}
\end{deluxetable}

\begin{deluxetable}{ccc}
\footnotesize
\tablecaption{Ulysses/SAX Annulus intersections with SAX WFC and LECS/MECS error circles}
\tablewidth{0pt}
\tablehead{
\colhead{Error circle} & \colhead{R.A. (2000)} & \colhead{Decl. (2000)} 
}
\startdata
SAX WFC  & 5\fh 01\fm 51\fs  & 11\arcdeg 43\arcmin 46\arcsec \nl
         & 5\fh 01\fm 45\fs  & 11\arcdeg 46\arcmin 02\arcsec \nl
         & 5\fh 01\fm 58\fs  & 11\arcdeg 43\arcmin 28\arcsec \nl
         & 5\fh 01\fm 45\fs  & 11\arcdeg 47\arcmin 34\arcsec \nl
SAX LECS/MECS & 5\fh 01\fm 45\fs  & 11\arcdeg 45\arcmin 58\arcsec \nl
         & 5\fh 01\fm 41\fs  & 11\arcdeg 47\arcmin 15\arcsec \nl
         & 5\fh 01\fm 47\fs  & 11\arcdeg 47\arcmin 01\arcsec \nl
         & 5\fh 01\fm 46\fs  & 11\arcdeg 47\arcmin 26\arcsec \nl
\enddata
\end{deluxetable}

\end{document}